\documentclass[journal,twocolumn]{IEEEtran}
\usepackage{epsfig,makeidx,color, multicol, here}
\usepackage{amsmath,amssymb,amsthm,bbm}
\usepackage{cite,graphicx}
\usepackage{enumerate}
\usepackage{hyperref}
\hypersetup{
	colorlinks = true,
	citecolor=blue,
}

\def\cA{{\cal A}}

\def\cQ{{\cal Q}}

\def\rT{{\rm T}}

\def\uP{{\mathbb P}}

\def\deft{ \buildrel \triangle \over = }

\def\be{ \begin{equation} }
\def\ee{ \end{equation} }
\def\bea{ \begin{eqnarray} }
\def\eea{ \end{eqnarray} }

\def\bs{{\bf s}}

\def\br{{\bf r}}

\def\bH{{\bf H}}

\def\bP{{\bf P}}

\def\b0{{\bf 0}}

\def\bpi{{\boldsymbol \pi}}

\ifCLASSOPTIONonecolumn
  \newcommand{\figwidth}{0.42\columnwidth}
  
\else
  \newcommand{\figwidth}{0.95\columnwidth}
\fi

\begin{document}

\title{Relay-aided Random Access for Energy-Limited Devices
in the IoT}

\author{Jinho Choi\\
\thanks{The author is with
the School of Information Technology,
Deakin University, Geelong, VIC 3220, Australia
(e-mail: jinho.choi@deakin.edu.au)}}

\date{today}
\maketitle

\begin{abstract}
In the Internet-of-Things (IoT), random access is employed for 
devices to share a common access channel in packet transmission with
low signaling overhead. Although a re-transmission strategy 
is necessary for packet collision resolution, it might be prohibitive for
some devices due to energy and complexity constraints.
In this paper, we consider 
a novel relay-aided random access (RARA)
scheme where re-transmissions are carried out
by relay nodes, not devices.
Thanks to multipacket reception with multiple copies of
collided signals forwarded by relay nodes,
a receiver is able to recover multiple collided packets simultaneously
in RARA.
As a result, devices of limited complexity and energy source 
can enjoy reliable transmission using RARA, and 
the throughput can approach 1
with a large number of relay nodes.
\end{abstract}

{\IEEEkeywords
Internet-of-Things (IoT);
Random Access; Relay Protocols;
Machine-Type Communications (MTC)}

\ifCLASSOPTIONonecolumn
\baselineskip 23.4pt
\fi

\section{Introduction} \label{S:Intro}

A large number of devices are to be connected 
in the Internet of Things (IoT) and
machine-type communication (MTC) has been considered
to support devices' connectivity for the IoT
in cellular systems \cite{Bockelmann16} \cite{3GPP_MTC} \cite{3GPP_NBIoT}.
Due to low signaling overhead, random access
is employed for MTC. For example, 
a multichannel ALOHA based random
access scheme is employed in \cite{3GPP_MTC}. 
In \cite{Arouk14} \cite{Choi16CL} \cite{Liu18},
various approaches to improve the performance of
multichannel ALOHA are considered to be used in MTC.


In random access, 
the devices that experience collisions can re-transmit collided
packets with a certain random backoff strategy \cite{Dai12}.
In this case, each device needs to implement a re-transmission scheme
with a buffer to keep packets before successful transmissions.
However, for devices with limited complexity and energy source, 
e.g., radio frequency identification (RFID) tags \cite{Dobkin12},
it is often difficult to use re-transmission schemes.

In this paper, we consider 
the case that devices
do not re-transmit collided packets in random access as they
have limited energy and complexity (and their
transmissions do not necessarily have to be reliable,
such as sensors' transmissions for environmental monitoring
with a large number of sensors).
This means that no random backoff strategy with buffer is employed at
devices. Instead,
we use relay nodes to re-transmit collided packets.
There have been various relay protocols for cooperative 
transmissions with relay nodes
\cite{Laneman04} \cite{Hossain11}.
While most cooperative communication approaches with relay are considered for
coordinated transmissions without contention, 
there are few approaches for random access, e.g., \cite{ElS12}.
Compared with the approach in \cite{ElS12},
the proposed approach in this paper can be easily
implemented at devices as they do not re-transmit the same packets.
Thus, it can be employed for MTC with devices 
of limited complexity and energy source.
However, in the proposed approach,
the transmission time can be long and some packets
may not be transmitted with a certain probability.
Thus, the approach
is not suitable for low-latency applications with high reliability. 
 

Note that unlike the approach in \cite{ElS12},
the proposed approach is based on
the multipacket reception \cite{Ghez88} 
\cite{Tsatsanis00} \cite{Bartoli17}
at a receiver to decode multiple collided
packets using multiuser detection (MUD) \cite{VerduBook},
which can result in a high throughput with a number of relay nodes.

\section{System Model} \label{S:SM}

Suppose that there are a number of IoT/MTC devices
and one base station (BS) that is a receiver
(i.e., we consider uplink transmissions from devices).
Each device becomes active when it has a data packet
to transmit and random access is considered to share
a given radio resource block.

\begin{figure}[thb]
\begin{center}
\includegraphics[width=\figwidth]{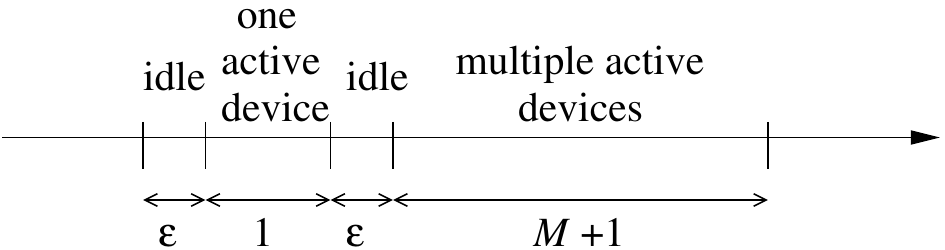}
\end{center}
\caption{Sessions of RARA (the 1st and 3rd sessions are idle,
the 2nd session has one active device, and the 4th session
has multiple active devices).}
        \label{Fig:rara}
\end{figure}

In this paper,
we assume that there are $M$ relay nodes
to help re-transmissions in random access.
The resulting scheme is referred to as relay-aided
random access (RARA).
In RARA, when active devices transmit 
their packets, we assume that the BS as well as $M$ relay nodes
receive the signals.
For convenience, we assume that the length of
packet is normalized.
As shown in Fig.~\ref{Fig:rara},
suppose that there is no active device in the first session.
Then, the BS can wait for a fraction time interval of $\epsilon \ (\le 1)$.
Since there is no signal, the BS can initiate the second session
(throughout the paper, it is assumed that the BS broadcasts a
signal to inform the devices the beginning of a new session).
In the second session, there is one active
device and there is no collision.
The BS can successfully receive the packet from the active device.
Subsequently, the BS broadcasts an acknowledgment (ACK) signal
at the end of the second session.
If there are multiple active devices as in the 4th session,
the BS fails to decode the packets due to packet collision and
broadcasts an negative-acknowledgment (NACK) signal.
In this case, $M$ relay nodes sequentially forward
their received signals to the BS. This requires $M$ unit times,
and the devices cannot
transmit their packets during the $M$ unit times.
It is assumed that the relay nodes are ordered in terms
of their transmissions in advance to avoid collisions.

For each active device, the session is 
the time interval (regardless of successful reception) 
at which the transmission is completed.
As shown in Fig.~\ref{Fig:rara}, the length of sessions can be
$\epsilon$ (for the state of idle), $1$ (for the state
of one active device), or $M+1$ (for the state
of multiple active devices). Note that
to shorten the idle time, $\epsilon$ can be small, i.e., $\epsilon \ll 1$.

Note that in \cite{ElS12}, a relay node is also considered
in random access to overcome channel fading
with spatial diversity. Thus, 
the motivation of making use of relay node
in \cite{ElS12} differs from that in this paper.
In addition, we consider multiple relay nodes and multipacket
reception (which are not considered in 
\cite{ElS12}) to decode multiple collided packets
without any complicated
temporal backoff strategies at devices and relay nodes.
As a result, the RARA scheme becomes
quite simple and suitable for MTC with devices of limited 
complexity and energy source.

\section{Multipacket Reception}

In this section, we discuss the multipacket reception 
\cite{Ghez88} \cite{Tsatsanis00} 
and show that
the BS is able to decode the collided signals from up to $M+1$ active devices
in RARA.

Suppose that there are $K$ active devices in a certain session,
say session $1$, where $K \ge 2$.
Let $h_k$ denote the channel coefficient
from the $k$th active device to the BS.
In addition, let $h_{m,k}$ and $g_m$ denote the channel coefficients
from the $k$th active device to the $m$th relay node and
from the $m$th relay node to the BS, respectively.
At the BS,
the received signal during session 1 is given by
\be
r_1 = \sum_{k=1}^K h_k s_k + n_1,
\ee
where $s_k$ is the packet transmitted from the $k$th active device
and $n_t$, $t = 1, \ldots$, is the background noise during session $t$.
According to RARA,
the $m$th relay node forwards its signal
(using the amplify-forward protocol \cite{Laneman04})
at session $m+1$ and the received signal at the BS becomes
\be
r_{m+1} = g_m x_m + n_{m+1}, \ m = 1, \ldots, M,
\ee
where $x_m$ is the signal transmitted by the $m$th relay node
that is given by
$x_m = \sum_{k=1}^K h_{m,k} s_k + w_m$.
Here, $w_m$ represents the background noise at the $m$th relay node.
In order to overcome packet
collisions and successfully decode the signals transmitted
by multiple active devices during session 1,
the BS can use the received signals in $M+1$ sessions,
i.e., $r_1, \ldots, r_{M+1}$. 

To illustrate multipacket reception in RARA,
for example, we consider the case of $M = 1$ (i.e., the case
that there is only one relay node).
Suppose that there are two active devices in a certain session 
(i.e., $K = 2$).
The BS fails to decode them due to the packet collision.
In the next session, the relay node forwards its received 
signal that is a superposition of the two signals from the two active devices.
The received signals from sessions 1 and 2 at the BS
can be written as
\begin{align}
\br = [r_1 \ r_2]^\rT 
= 
\bH \bs + 
\left[
\begin{array}{cc}
n_1 \cr
g_1 w_1 + n_2 \cr
\end{array}
\right] ,
	\label{EQ:22}
\end{align}
where $\bs = [s_1 \ s_2]^\rT$, $\bH=
\left[ \begin{array}{cc}
h_1 & h_2 \cr
g_1 h_{1,1} & g_2 h_{1,2} \cr
\end{array} \right]$,
and the superscript $\rT$
represents the transpose.
If the rank of $\bH$ is 2,
$\bs$ can be recovered by 
multiplying $\br$ by $\bH^{-1}$ (ignoring the noise terms).
The resulting detector is called the decorrelating detector
\cite{VerduBook}.
We can generalize it to large $K$ and $M$ where
the size of the matrix $\bH$ becomes 
$(M+1) \times K$. As long as the rank
of $\bH$ is $K$
(with $M +1 \ge K$), the inverse or pseudo-inverse
$\bH$ can be used to find $\bs$ for multipacket reception.
Other MUD algorithms \cite{VerduBook} can be 
employed to recover $\bs$ from $\br$.
From this, with a sufficiently high signal-to-noise ratio 
(SNR), we can see that the BS can decode up to $M+1$ 
collided signals in RARA.
Consequently, throughout the paper, we assume\footnote{The
performance of multipacket reception 
in RARA depends on the channel coefficients, $\{g_m\}$, $\{h_k\}$,
and $\{h_{m,k}\}$. However, for a tractable link-layer
performance analysis, we simply assume that multipacket reception 
becomes successful if $K \le M+1$. In the future,
for a more realistic performance analysis,
the performance of multipacket reception is to be taken into
account.}
that the BS is able to decode the signals from up to $M+1$ active
devices.  Here, $M$ becomes a design parameter.


\section{Analysis}	\label{S:Anal}

For performance analysis, we consider
the following four states for each session:
\emph{i)} State $0$: idle, i.e.,
no active device;
\emph{ii)} State $1$: one active device;
\emph{iii)} State $S$: $k$ active devices and
successful decoding, where $k \in \{2,\ldots, M+1\}$;
\emph{iv)} State $U$: $k$ active devices and unsuccessful
decoding, where $k \ge M+2$.
For convenience, let $\cA = \{0, 1, S, U\}$ denote
the set of states.
Furthermore, 
we assume a Poisson distribution to model the number of active devices.
In particular, the probability of $k$ active devices
over $T$ unit times is given by
\be
\Pr(k\,|\, T) = e^{-\lambda T} \frac{(\lambda T)^k }{k!},
	\label{EQ:PoiD}
\ee
where $\lambda$ is the traffic intensity per unit time
(i.e., the average number of active devices per unit time is $\lambda$).
In addition, for convenience, let
$p_{0|T} = \Pr(0\,|\, T)$,
$p_{1|T} = \Pr(1\,|\, T)$,
$p_{S|T} = \Pr(k \in \{2,\ldots, M+1\}\,|\, T)$,
and $p_{U|T} = \Pr(k \ge M+2\,|\, T)$.
Since the state of a session
depends on the state of the previous session, 
we have a Markov chain with
the following state transition matrix:
\be
\bP = \left[
\begin{array}{cccc}
p_{0|\epsilon} & p_{1|\epsilon} & p_{S|\epsilon} & p_{U|\epsilon} \cr
p_{0|1} & p_{1|1} & p_{S|1} & p_{U|1} \cr
p_{0|M+1} & p_{1|M+1} & p_{S|M+1} & p_{U|M+1} \cr
p_{0|M+1} & p_{1|M+1} & p_{S|M+1} & p_{U|M+1} \cr
\end{array}
\right],
	\label{EQ:Pt}
\ee
where $[\bP]_{i,j} =
P_{i,j} = \Pr(x_{t+1} = j \,|\, x_t = i)$, $i,j \in \cA$.

Denote by $\pi_i$, $i \in \cA$, the stationary distribution,
and let $\bpi = [\pi_0 \ \pi_1 \ \pi_S \ \pi_U]$,
where $\pi_i$ represents the probability that
the state of a session is $i \in \cA$ in steady-state.
Since
$\bpi = \bpi \bP$, after some manipulations, we can have
\begin{align}
\pi_0 & = (p_{0|M+1} (1- p_{1|1}) + p_{1|M+1} p_{0|1})/\Delta \cr
\pi_1 & = (p_{1|M+1} (1- p_{0|\epsilon}) + p_{0|M+1}p_{1|\epsilon})/\Delta\cr
\bar \pi & = 
( (1 - p_{1|1})(1-p_{0|\epsilon}) - p_{0|1} p_{1|\epsilon})/\Delta,
	\label{EQ:pis}
\end{align}
where $\bar \pi = \pi_S + \pi_U$
and
\begin{align*}
\Delta 
& = (p_{0|M+1} + 1 - p_{0|\epsilon})(p_{1|M+1} + 1 - p_{1|1}) \cr
& \quad - (p_{0|M+1} -p_{0|1})(p_{1|M+1} - p_{1|\epsilon}).
\end{align*}
Note that once $\pi_0$, $\pi_1$, and $\bar \pi$ 
are found, we can obtain $\pi_S$ and $\pi_U$ as follows:
$$
\pi_i = p_{i|\epsilon} \pi_0 + 
p_{i|1} \pi_1 + p_{i|M+1} \bar \pi, \ i \in \{S,U\}.
$$

For a large $M$, if there are two or more active devices 
in a session, 
the state of the next session 
will be $S$ or $U$ with a high probability since the session time is long.
In particular,
according to \eqref{EQ:PoiD} and \eqref{EQ:pis},
we have $p_{0|M+1}, p_{1|M+1} \to 0$ and $\bar \pi \to 1$
as $M \to \infty$. Furthermore, as $M \to \infty$, we have 
\be
\pi_S \approx p_{S|M+1} \ \mbox{and} \ 
\pi_U \approx p_{U|M+1}.
\ee

\subsection{Outage Probability and Throughput}

Let $Q(k)$ be the probability that
there are $k$ active devices in a session.
Using the stationary distribution,
it can be shown that
\be
Q(k) = \sum_{i \in \cA} \Pr(k\,|\, T_i) \pi_i,
\ee
where 
$T_i = \epsilon$, $1$, and $M +1$ if 
$i = 0$, $1$, and $i \in \{S,U\}$, respectively.
If the number of active devices
is greater than or equal to $M+2$ in a session, 
the BS cannot decode the signals although it receives
the forwarded signals from $M$ relay nodes. The corresponding
event is referred to as the outage event and
the outage probability is given by
\begin{align}
\uP_{\rm out} (\lambda, M) 
= \sum_{k=M+2}^\infty Q(k) 
= \sum_{i \in \cA}  \sum_{k=M+2}^\infty
e^{-\lambda T_i} \frac{(\lambda T_i)^k}{k!} \pi_i,
	\label{EQ:Pout}
\end{align}
which is a function of $\lambda$ and $M$.
In general, it is expected to have a low outage probability.

The throughput can be defined as the ratio
of the average number of successfully transmitted
packets (per session) to the average length of session.
The average length of session is given by
\begin{align}
\bar T = \epsilon \pi_0 + \pi_1 + (M+1) \bar \pi 
\le 1 + M \bar \pi.
	\label{EQ:bT}
\end{align}
The
average number of successfully transmitted
packets (per session) 
can be found as
\begin{align}
\bar K & = \sum_{k=1}^{M+1} k Q(k) 
= \sum_{i \in \cA} \sum_{k=1}^{M+1} k \Pr (k\,|\, T_i) \pi_i \cr
& = \sum_{i \in \cA} 
\left(\sum_{k=0}^M \frac{(\lambda T_i)^k}{k!} 
\right) e^{- \lambda T_i}  \lambda T_i \pi_i .
	\label{EQ:bK}
\end{align}
The throughput becomes
\be
\eta (\lambda, M) = \frac{\bar K}{\bar T},
	\label{EQ:elM}
\ee
which is neither an increasing nor decreasing function of $M$
(as can be shown in Fig.~\ref{Fig:plt1} (a)).

\subsection{Asymptotic Results with $M \to \infty$}

Suppose that $\lambda$ is fixed, while $M \to \infty$. 
In this case, from \cite{MacKayBook}, 
we can have the following Gaussian approximation:
\begin{align}
& \lim_{M \to \infty}
\sum_{k=0}^M \frac{(\lambda (M+1))^k}{k!} e^{-\lambda (M+1)}\cr
& = 
\lim_{M \to \infty}
F_{\rm P} (M;\lambda (M+1)) \cr
& = 
\lim_{M \to \infty}
F_{\rm G} (M; \lambda (M+1), \lambda (M+1)),
	\label{EQ:GA}
\end{align}
where $F_{\rm P} (x; \lambda)$
denotes
the cumulative distribution function
(cdf) of a Poisson random variable with parameter $\lambda$
and
$F_{\rm G}(x; \mu, \sigma^2)$ represents
the cdf of a Gaussian random variable
with mean $\mu$ and variance $\sigma^2$.
Since
\begin{align}
F_{\rm G} (M; \lambda (M+1), \lambda (M+1)) 
=1-\cQ \left( \psi(\lambda, M) \right),
\end{align}
where $\psi(\lambda, M) = (1 - \lambda)
\sqrt{\frac{M}{\lambda}}$ and
$\cQ (x) = \int_x^\infty
\frac{e^{-\frac{t^2}{2}}}{\sqrt{2 \pi}} dt$ is the Q-function,
we have
\begin{align}
\lim_{M \to \infty}
\sum_{k=0}^M \frac{(\lambda (M+1))^k}{k!} e^{-\lambda (M+1)}
= U(\lambda) \deft \left\{
\begin{array}{ll}
1, & \mbox{if $\lambda < 1$} \cr
\frac{1}{2}, & \mbox{if $\lambda = 1$} \cr
0, & \mbox{if $\lambda > 1$.} \cr
\end{array}
\right.
	\label{EQ:Ul}
\end{align}
In addition, since $\bar \pi \to 1$ as $M \to \infty$,
by substituting \eqref{EQ:Ul} into \eqref{EQ:bK},
it can be shown that
the asymptotic throughput is given by
\be
\lim_{M \to \infty} \eta (\lambda, M)  
= \frac{\lambda (M+1) \bar \pi U(\lambda)}{(M+1) \bar \pi}=
\lambda U(\lambda).
	\label{EQ:AT}
\ee
The result in \eqref{EQ:AT} demonstrates
that the throughput of RARA can approach 1
(with a large $M$ and $\lambda = 1 - \delta$
with a sufficiently small $\delta > 0$).
Since the maximum throughput of slotted ALOHA is $e^{-1}$ 
\cite{BertsekasBook},
we can have a throughput improvement by a factor
of $e \approx 2.718$ thanks to 
multipacket reception.

Using the Gaussian approximation in 
\eqref{EQ:GA}, for a sufficiently large $M$, we have
\be
\eta (\lambda, M) \approx \lambda 
\left(1 - \cQ (\psi(\lambda, M)) \right),
	\label{EQ:ape}
\ee
which allows us to find the optimal value of $\lambda$
that maximizes the throughput for a given large $M$.
Since $\cQ (\psi(\lambda, M))$ is an increasing function of $\lambda$
(for a fixed $M$), we may expect that the throughput
increases and then decreases with $\lambda$.
Similarly, for a large $M$, the outage probability can be approximated as
\be
\uP_{\rm out} (\lambda, M) \approx 
 \cQ (\psi(\lambda, M)),
	\label{EQ:apP}
\ee
which shows that $\lambda$ has to be less
than $1$ for a low outage probability, which decreases with $M$.
In addition, it can be shown that for a fixed $\lambda$, 
$\lim_{M \to \infty} \uP_{\rm out} (\lambda, M) 
= 1 - U(\lambda)$.

\section{Simulation Results}

In this section, we present simulation results under
the assumptions in Section~\ref{S:Anal} except that
$K$ is finite ($K$ is a binomial random variable $n = 40$ and
$p = n/\lambda$).

\begin{figure}[thb]
\begin{center}
\includegraphics[width=\figwidth]{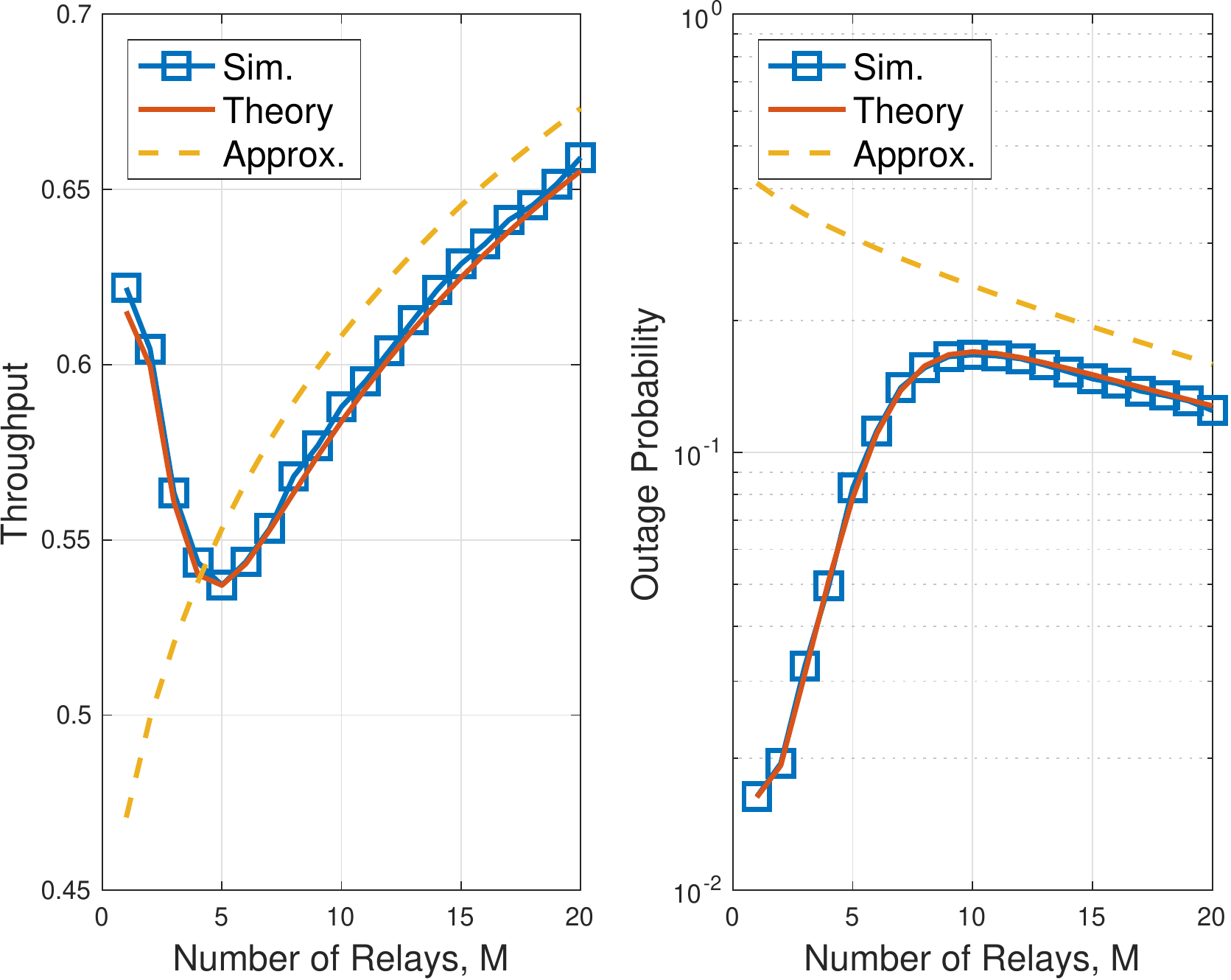} \\
\hskip 0.5cm (a) \hskip 3.5cm (b) 
\end{center}
\caption{Performance
of the RARA scheme for different numbers of relay nodes, $M$
when $\lambda = 0.8$ and $\epsilon = 0.1$:
(a) throughput; (b) outage probability.}
        \label{Fig:plt1}
\end{figure}

In Fig.~\ref{Fig:plt1},
we show the throughput and outage probability
as functions of the number of relay nodes, $M$,
when $\lambda = 0.8$ and $\epsilon = 0.1$.
We assume that the number of devices is $40 \times M$
and each device becomes active 
(per unit time) with probability $p_a = \frac{\lambda}{40 M}$.
Since the number of devices is large,
although the number of devices is a binomial random variable,
it can be approximated by a Poisson random variable
as in \eqref{EQ:PoiD}.
We can see that the theoretical throughput
and the outage probability in \eqref{EQ:elM} 
and \eqref{EQ:Pout}, respectively,
agree with the simulation results,
while the approximations in \eqref{EQ:ape}
and \eqref{EQ:apP} are also 
reasonably tight when $M$ is sufficiently large.

From Fig.~\ref{Fig:plt1} (a), we see that
the throughput decreases and then increases with $M$. 
Thus, it might be necessary to use one relay node ($M = 1$)
or a number of relay nodes ($M \ge 20$) to achieve
a good throughput with a low outage probability
as in Fig.~\ref{Fig:plt1} (b).

\begin{figure}[thb]
\begin{center}
\includegraphics[width=\figwidth]{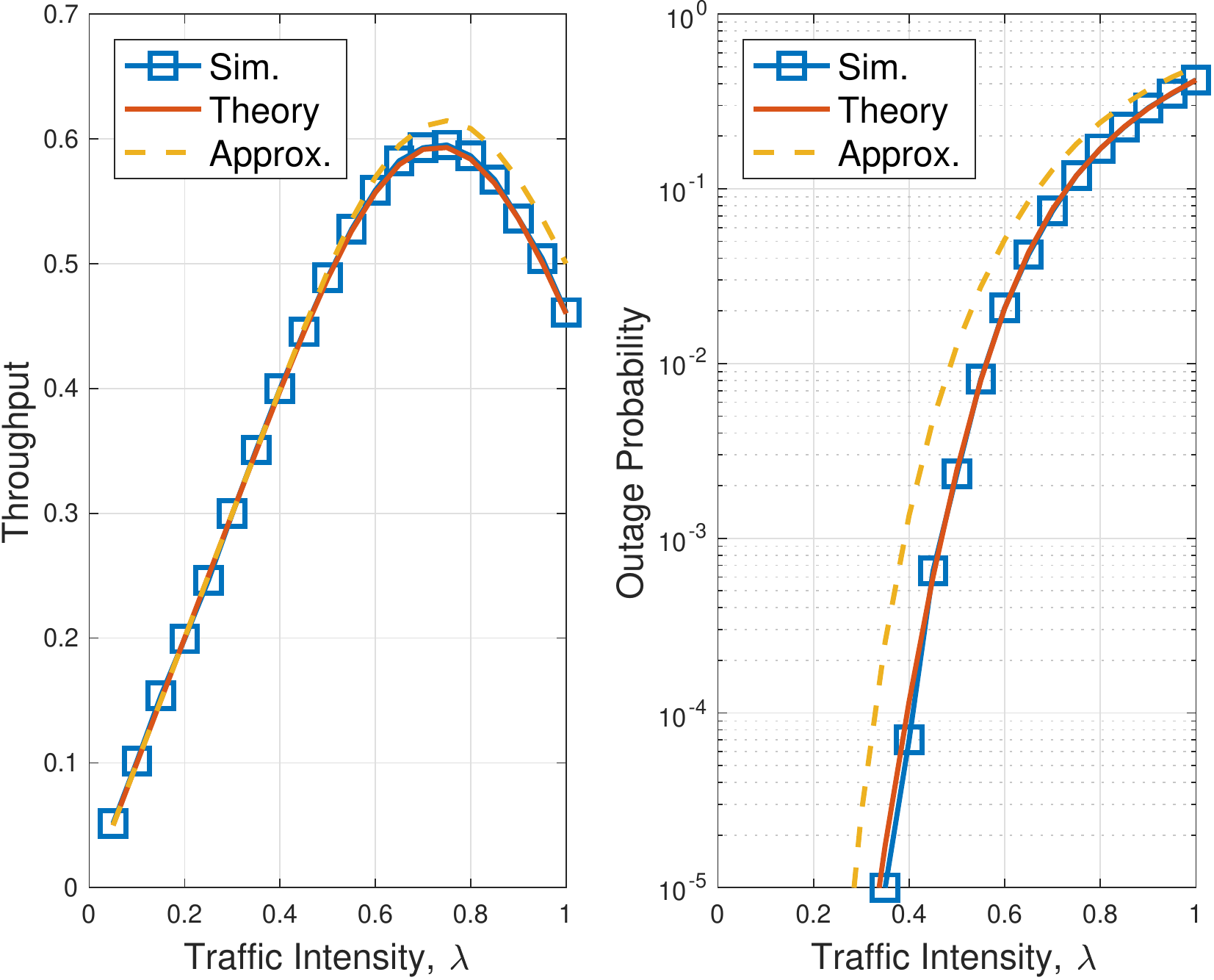} \\
\hskip 0.5cm (a) \hskip 3.5cm (b) 
\end{center}
\caption{Performance
of the RARA scheme for different values of $\lambda$
when $M = 10$ and $\epsilon = 0.1$:
(a) throughput; (b) outage probability.}
        \label{Fig:plt2}
\end{figure}

Fig.~\ref{Fig:plt2} shows
the throughput and outage probability
as functions of $\lambda$
when $M = 10$ and $\epsilon = 0.1$.
It is shown that the throughput has a peak at around $\lambda = 0.7$
and the outage probability increases with $\lambda$,
which is expected by \eqref{EQ:apP}.

\section{Conclusions}

In this paper, we proposed a random access scheme,
i.e., the RARA scheme, that does not require re-transmissions by devices,
but relay nodes for packet collision resolution in MTC.
Thanks to relay nodes, it was possible for devices to reliably transmit 
their packets in random access without employing complicated re-transmission
schemes. Thus, the proposed approach can be used for
devices of limited complexity and energy source 
(e.g., RFID tags) in the IoT.

\bibliographystyle{ieeetr}
\bibliography{mtc}

\end{document}